\documentclass[aps,prl,reprint,groupedaddress]{revtex4-2}

\usepackage{amsmath}
\usepackage{amssymb}
\usepackage{bm}
\usepackage{amsmath,amssymb,multirow,graphicx}
\usepackage{subfigure}
\usepackage{epstopdf}
\usepackage{subcaption}
\usepackage{cancel}
\usepackage{tikz}
\usepackage{graphicx}

\newtheorem{myconj}{Conjecture}

\usepackage{comment}
\usepackage{color}%

\def\hhref#1{\href{http://arxiv.org/abs/#1}{arXiv:#1}} 

\graphicspath{{./Figures/}}

 \usepackage{stackrel,nicefrac}
 \usepackage{mathtools}

\usepackage{hyperref}
 
\begin{document}

\title
{Exact effective central charges for a family of $3d$ superconformal field theories}

\author{Griffen Adams}
\affiliation{Department of Physics, University of Connecticut,
Storrs, CT 06269-3046}

\begin{abstract}
We conjecture an exact formula for the effective central charge, $c_{\rm eff}$, of a family of three-dimensional superconformal field theories related to Chern-Simons theory on a three-manifold by the $3d$-$3d$ correspondence. The formula determines $c_{\rm eff}$, which governs the growth of a protected sector of states in the conformal theory, in terms of the dimension of a particular space of modular forms. We confirm our approach by checking against known results computed through resurgent continuation.
\end{abstract}

\maketitle

Dualities in quantum field theory have been an increasingly important tool for studying non-perturbative effects in physics in the past few decades, particularly in low-dimensional QFTs. The $3d$-$3d$ correspondence is an example of these dualities which relates complex Chern-Simons theory on a three-manifold $M_3$ to a three-dimensional $\mathcal{N}=2$ superconformal field theory living on a simple three-dimensional space, but whose physical content is labelled by $M_3$ \cite{DGG1,DGG2,CGDS3d3d}. The $\mathcal{N}=2$ theory, denoted $T[M_3]$, is intrinsically a strongly coupled theory and does not generically have a known Lagrangian description which makes computations difficult \cite{GPV,GPPV}. However, by computing certain quantities in Chern-Simons theory we can answer many questions on the supersymmetric side. An important prediction of the correspondence is the equality of the $T[M_3]$ partition function on $D^2 \times S^1$, known as the half-index, and the quantum $\widehat{Z}$ invariants of Chern-Simons theory, which are $q$-series with integer coefficients defined at $|q|<1$ \cite{GPV,GPPV}. A more natural quantity is the superconformal index, or the partition function on $S^2 \times S^1$, which has a more complicated interpretation on the Chern-Simons side:
\begin{eqnarray}
    \mathcal{I}(q) = \frac{1}{\eta(q)^2}\sum_a |\mathcal{W}_a| \widehat{Z}_a(q)\widehat{Z}_a(q^{-1})
    \label{eq:SCFTindex}
\end{eqnarray}
Here $\widehat{Z}_a(q^{-1})$ is the extension of $\widehat{Z}_a(q)$ to the other side of the unit circle, $|q|>1$, $a$ labels elements of ${\rm Spin^c}(M_3)$, and the $\mathcal{W}_a$ are symmetry factors \cite{GPPV}. In most known cases, $\widehat{Z}_a(q)$ has a dense set of singularities on the unit circle, forming a natural boundary to analytic continuation \cite{3dmod}. There have been several proposals for how to construct the continuation of $\widehat{Z}_a(q)$ across this boundary \cite{GMP16,CDGG23,Park21,3dmodre,ACDGO,MJ26}. In Chern-Simons theory, the variable $q$ is the exponential of the negative coupling, $q=e^{-\hbar}$, and thus a change in sign of the coupling implements the continuation $q \to q^{-1}$, crossing the boundary. The action functional of Chern-Simons theory provides a clear physical interpretation of this continuation as an {\it orientation reversal} $M_3 \to \overline{M_3}$ of the manifold, identifying $\widehat{Z}_a(M_3;q^{-1})$ with $\widehat{Z}_a(\overline{M_3};q)$ \cite{GPPV,3dmod}. In general this is a new $q$-series that is dual to $\widehat{Z}(M_3;q)$ under $q \to q^{-1}$ with different coefficients and asymptotics, and its construction is a strong test of the Chern-Simons path integral and its continuation. A method to compute these $q$-series expansions was developed recently using ideas from resurgence analysis, allowing for access to the full superconformal index \cite{CDGG23,ACDGO}.

One of the most fundamental quantities in a conformal field theory is the central charge, which in two dimensional theories plays two roles: it fixes the symmetry algebra governing the quantum theory, and controls the growth of the spectrum through Cardy's formula, $\log b_n \sim n^{1/2}$ \cite{Cardy2d}. For three dimensional theories we lose much of this, but the second role admits a natural generalization through the asymptotics of the superconformal index. While the number of states in a generic $3d$ CFT grows as $\log b_n \sim n^{2/3}$, the sector counted via the index has the same Cardy formula-like asymptotics as $2d$ theories \cite{GJ23}. This allows us to define the {\it effective central charge} $c_{\rm eff}$,
\begin{eqnarray}
    \mathcal{I}(q) = \sum_n b_n q^n, \quad &\log& |b_n| \sim \sqrt{\frac{2\pi^2}{3}c^{\mathcal{I}}_{\rm eff} \,n}
    \label{eq:indexgrowth}
\end{eqnarray}
This can be expressed in terms of the asymptotics of the half indices $\widehat{Z}_a(M_3;q)$ and $\widehat{Z}_a(\overline{M_3};q)$:
\begin{eqnarray}
    c_{\rm eff}^{\mathcal{I}} = 2+c_{\rm eff}^{T[M_3]} + c_{\rm eff}^{T[\overline{M_3}]}
\end{eqnarray}
where the factors of $\eta(q)$ contribute the $+2$ to the full effective central charge, and the second two terms are defined via the $q$-series expansions of the half indices $\widehat{Z}_a(q)$ and $\widehat{Z}_a(q^{-1})$, analogously to \eqref{eq:indexgrowth}. Note that this normalization of differs from that of \cite{ACDGO2}, where $c_{\rm eff}^{T[M_3]}$ and $c_{\rm eff}^{T[\overline{M_3}]}$ both included a contribution of $+1$ from one factor of $\eta(q)$ each.

In this paper, we focus on the case where $M_3$ is a Seifert homology sphere with three exceptional fibers, denoted $\Sigma(p_1,p_2,p_3)$, with mutually coprime $p_i$ and a unique ${\rm Spin^c}$ structure, whose label we omit. We will also refer to these manifolds as Brieskorn spheres. The $\widehat{Z}(M_3;q)$ are expressed in terms of weight $1/2$ false theta functions with modular completions constructed from weight $3/2$ unary theta functions, which are associated to Weil representations $\Theta^{m+K}$ of the double cover of $SL(2,\mathbb{Z})$ \cite{GPPV,3dmod,3dmodre,LawrenceZagier,Hikami04}. The construction of these representations is described in detail in \cite{3dmod,3dmodre}, and we refer the reader to them for a full exposition. There exists a one-to-one map between the Seifert labels $(p_1,p_2,p_3)$ and the Weil representations considered in this work
\begin{eqnarray}
    m=p_1p_2p_3, \quad K=\{p_1p_2,p_1p_3,p_2p_3\}
\end{eqnarray}
and we will use both labels interchangeably. False theta functions have $q$-series coefficients $a_n \in \{-1,0,1\}$, and thus do not contribute to the $\log b_n$ asymptotics, which gives $c_{\rm eff}^{T[M_3]}=0$.

The difficult part of computing $c_{\rm eff}^{\mathcal{I}}$ comes from $\widehat{Z}(\overline{M_3};q)$, the extension of $\widehat{Z}(M_3;q)$ across the unit circle, whose coefficients have asymptotic growth of the form
\begin{eqnarray}
    \widehat{Z}^{[j]}(\overline{M_3};q) &=& q^{-\Delta_j} \sum_{n} a_{n,j} q^{n}, \\ a_{n,j} &\sim&  \exp\left(\sqrt{\frac{2\pi^2}{3} c_{\rm eff}^{T[\overline{M_3}]}\, n}\right)
    \label{eq:zhatgrowth}
\end{eqnarray}
where $\Delta_j=r_j^2/4m \pmod{1}$ are the Chern-Simons values of $M_3$ ordered from smallest to largest in the interval $(0,1)$ and are determined by the representation $\Theta^{m+K}$, or equivalently the Seifert labels $(p_1,p_2,p_3)$. Note that the $\widehat{Z}$-invariant is packaged as a vector - a physical interpretation of the other entries as partition functions of $T[M_3]$ with defects inserted was given in \cite{3dmodre}. When $M_3$ is a Seifert homology sphere, the orientation reversed invariants are vector-valued mock modular forms, which have appeared in various areas of physics. We will use the notation $c_{\rm eff} := c_{\rm eff}^{T[\overline{M_3}]}$, since the orientation reversed side will be the focus of this letter. This quantity is completely determined by one of the Chern-Simons values, which in practice is the largest $\Delta_j$ such that $a_{0,j} \neq 0$, or the most polar term in the $q$-series vector \cite{ACDGO2}. We denote this distinguished value by $\widetilde{\Delta}$, and with our particular normalization we have 
\begin{eqnarray}
    c_{\rm eff} = 24\widetilde{\Delta}
\end{eqnarray}
There has been recent success in determining this quantity for the homology spheres at small values $(p_1,p_2,p_3)$. This was initiated in \cite{GJ23}, where Park's surgery formula was used to study the large order asymptotics of the $\widehat{Z}$ invariants \cite{Park21,GM19}, and was recently extended to a much larger class of manifolds \cite{ceffmod25}. Another recent method, which produces different answers for $c_{\rm eff}$, relies on principles from resurgence, and determines $c_{\rm eff}$ by finding the polar terms of the vector of $\widehat{Z}$ invariants for the orientation reversed manifolds \cite{ACDGO2}. The goal of this work is to find a more fundamental mechanism underlying this effective central charge that will allow us to explore large values of $(p_1,p_2,p_3)$.

Let $c_{\rm eff}(p_1,p_2,p_3)$ be the effective central charge associated to the orientation reversed manifold $\overline{\Sigma(p_1,p_2,p_3)}$. Below we display some results obtained using resurgence in \cite{ACDGO2},
\begin{eqnarray}
    c_{\rm eff}(2,3,5) &=& 1/5, \,\,\,\,\,\,\quad c_{\rm eff}(2,3,7) = 1/7, \\ c_{\rm eff}(2,3,11)&=& 25/11, \quad c_{\rm eff}(2,3,13) = 1/13 \nonumber
\end{eqnarray}
In the vector of $q$-series for $\overline{\Sigma(2,3,11)}$ there are two polar terms, $q^{-1/264}$ and $q^{-25/264}$, the latter of which fixes $c_{\rm eff}$. This is in contrast to the three other examples listed, which are referred to as ``optimal" in the language of mock Jacobi forms and umbral moonshine, and only have a single polar term, $q^{-1/4m}$ \cite{DMZ12,CD16opt,CDHumbral}. The optimal examples are attached to representations $\Theta^{m+K}$ whose corresponding modular group $\Gamma_0(m) + K$ has genus zero \cite{CD16opt,Fer93}. However, unlike in these optimal examples, there is no known mechanism that tells us that the representation associated to $(2,3,11)$ forces $\Delta_2=25/264$ to be the deepest pole in the $\widehat{Z}$ vector, and similarly for the other non-optimal examples. There is a finite list of $(p_1,p_2,p_3)$ that are optimal, and it would be useful to determine such a mechanism that gives $c_{\rm eff}$ purely from representation theory data. It is reasonable to ask if there exists an object associated to $\Theta^{m+K}$ with an integer invariant $N$ that picks out $\widetilde{\Delta}=\Delta_N$, thereby giving $c_{\rm eff}$.

In the following section, we argue that this object is the space of weight $3/2$ cusp forms, denoted $S_{3/2}(\Theta^{m+K})$, and show that its dimension $N$ can be calculated through a Riemann-Roch type formula \eqref{eq:rrN}. Then, we confirm that this agrees with the known results for $c_{\rm eff}$ found by resurgent continuation, and study the behavior of $c_{\rm eff}$ as the index $m=p_1p_2p_3$ gets large. From the observed behavior, we conjecture simple and exact formulas for $c_{\rm eff}$ when the labels satisfy certain number theoretic congruences. Finally, we comment on the case of Seifert manifolds with four exceptional fibers, and conclude by giving explicit computations for examples with three and four fibers.

First, we review how the dual $q$-series for the orientation reversed manifolds $\overline{M_3}$, $\widehat{Z}^{[j]}(\overline{M_3};q)$, are constructed using resurgent continuation \cite{ACDGO}. Given a three-fibered Seifert homology sphere $\Sigma(p_1,p_2,p_3)$, there is a dimension $d$ vector of false theta functions $\Psi^{m+K}(q)$ giving the half index and their defect partners, where $d$ is given by
\begin{eqnarray}
    d= \frac{1}{4}(p_1-1)(p_2-1)(p_3-1).
\end{eqnarray}
for prime $p_i$, which will be the class that we restrict to in this letter. These are attached to the representation $\Theta^{m+K}$, which has $\mathcal{S}$ and $\mathcal{T}$ generators. Under the $\mathcal{S}$ transformation, which takes $q \to \tilde{q}=e^{-\pi^2/\hbar}$, the vector of false theta functions transform as \cite{BringmannFalse,ACDGO}
\begin{eqnarray}
    \Psi^{m+K}(q) + i\sqrt\frac{\pi}{|\hbar|} S^{m+K}\Psi^{m+K}(\tilde{q}) = J(\hbar<0)
\end{eqnarray}
where $J(\hbar<0)$ is the vector of Eichler integrals of the weight $3/2$ unary theta functions $\theta^{m+K}$ at $\hbar<0$, and $S^{m+K}$ is the matrix implementing the $S$-transformation in the $\Theta^{m+K}$ representation. This expression can be viewed as the unique transseries representation of $J(\hbar)$ on the Stokes line $\hbar<0$ \cite{ACDGO,ACDGO2}. The integrals can be rotated to $\hbar>0$, and the property of preservation of relations of resurgent functions fixes the transseries structure of the $q/\tilde{q}$ split,
\begin{eqnarray}
    X^{m+K}(q) + \sqrt\frac{\pi}{|\hbar|} S^{m+K}X^{m+K}(\tilde{q}) = J(\hbar>0)
    \label{eq:Sdual}
\end{eqnarray}
This produces a vector of mock modular forms $X^{m+K}(q) :=\widehat{Z}(\overline{\Sigma(p_1,p_2,p_3)};q)$ with the same unary theta shadow as the false theta functions. The uniqueness and existence properties of $X^{m+K}$ have recently been discussed \cite{ACDGO,CDS,CDS2}. There are solutions to the homogeneous equation, where $J(\hbar)=0$, that can be added to the solutions $X^{m+K}(q)$ and change their asymptotics, but there exists a representative modulo these homogeneous solutions with a minimal growth rate, which appears to be the exact vector of $q$-series computed by resurgent continuation.

First, we argue that this minimal growth rate is related to the number of constraints imposed by the Bruinier-Funke pairing. The mock modular forms $X^{m+K}$ fit very naturally in the theory of weak Maass forms, functions $f$ with canonical holomorphic/non-holomorphic decompositions $f=f^+ + f^-$ among other properties \cite{Zwegers08,Bringmann17,BFlifts04}. The $f^+$ are the mock modular forms $X^{m+K}$, while the $f^-$ are their completions, satisfying $\xi_{1/2} f^- = \theta^{m+K}$, where $\xi_{k}$ is the weight $k$ modular raising/lowering operator. Then, fixing a basis of cusp forms $g_i \in S_{3/2}({\Theta^{m+K}})$, a theorem of Bruinier and Funke gives \cite{BFlifts04,KaneMan17}
\begin{eqnarray}
    \sum_{j=1}^d\sum_{n>0} a_{n,j}^{g_i} a_{-n,j}^X = (g_i,\theta^{m+K})_{3/2}
\end{eqnarray}
where $a_n^g$ and $a_n^X$ refer to the $q$ coefficients of $g_i$ and $X^{m+K}$ respectively. Here, $(f,g)_{3/2}$ is the weight $3/2$ Petersson scalar product, which for fixed $f$ gives a map from $S_{3/2}({\Theta^{m+K}})$ to $\mathbb{C}$, and thus a linear $N \times M$ system with $N=\dim S_{3/2}({\Theta^{m+K}})$ and $M$ the number of polar coefficients in $X^{m+K}$. 

This gives $N$ linear conditions on the $M$ polar coefficients of $X^{m+K}$. For generic $g_i$ existence of a solution requires $N \leq M$, and recent work suggests that the dual $q$-series satisfy a minimal growth condition among the solutions to the $S$-transformation equation \eqref{eq:Sdual}, implying that $M$ saturates this bound and $N=M$. Assuming further that this minimal solution $X^{m+K}$ contains precisely the $N$ shallowest polar terms $\Delta_1,\dots,\Delta_N$, we find that $\Delta_N$ determines the leading exponential growth and thus $c_{\rm eff}$. We will see that this is true for all examples given in \cite{ACDGO2}, see Table \ref{tab:tab1}.

\begin{figure}
    \centering
    \includegraphics[width=1\linewidth]{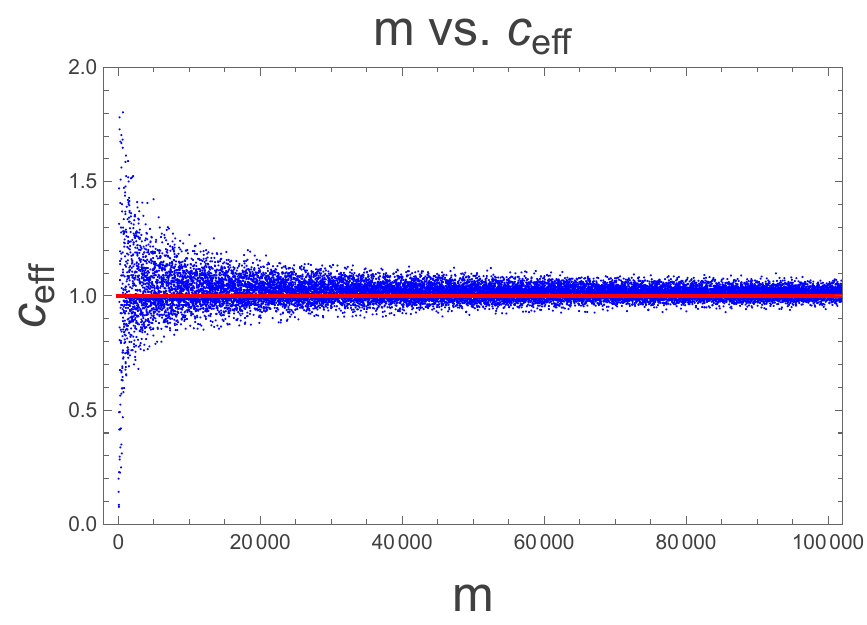}
    \caption{Values of $c_{\rm eff}$ as a function of $m=p_1p_2p_3$ (blue dots) for $m<10^5$. We only consider prime $p_i$ so that each $m$ is a unique representation $\Theta^{m+K}$. A red line at $c_{\rm eff}=1$ is also plotted.}
    \label{fig:allP}
\end{figure}

Formulae for computing dimensions of spaces of vector-valued modular forms have been studied extensively \cite{Gannon13,CFK17}, and have been utilized in other physical contexts where modularity plays a role \cite{SA24}. In particular, a formula for the dimension of the cusp form space associated to representations of the double cover of $SL(2,\mathbb{Z})$, $\rho$, can be derived from an application of the Riemann-Roch theorem, proven in \cite{Gannon13},
\begin{eqnarray}
    \dim \mathcal{M}_k(\rho) &-& \dim \mathcal{M}_{2-k}(\bar\rho) = c_{(\rho,k)} + d- {\rm Tr} \lambda     \label{eq:gannonRR} \\ \nonumber \\
    c_{(\rho,k)} &=& \frac{kd}{12}-\frac{\alpha_1}{2}-\frac{\beta_1+2\beta_2}{3} \nonumber
\end{eqnarray}
In this formula, $\mathcal{M}_k$ is the space of vector-valued modular forms of weight $k$, $d$ is the dimension of the representation $\rho$, and $\lambda = {\rm diag}(\Delta_1, \Delta_2, \dots \Delta_d)$ is just the diagonal matrix of $T$-eigenvalues. The $\alpha_n$ are the multiplicities of $(-1)^n$ as eigenvalues of the $S$ matrix, and the $\beta_n$ are the multiplicities of $\zeta_3^n = e^{2\pi in/3}$ as eigenvalues of the matrix $\mu ST$, where $\mu$ is chosen such that $(\mu S T)^3=I$. Some further simplifications of \eqref{eq:gannonRR} can be made: for the representations $\rho = \Theta^{m+K}$, we have $\Delta_j \notin \mathbb{Z}$, so the spaces of holomorphic forms $\mathcal{M}_k$ and cusp forms $\mathcal{S}_k$ have the same dimension and are interchangeable. It can also be shown that when $k=3/2$, $\mathcal{S}_{2-k}(\bar\rho)$ vanishes if $\rho$ is a Weil representation of index $m$, as it embeds in the space of weight $1$ Jacobi forms $J_{1,m}$, which was shown to vanish by a theorem of Skoruppa \cite{Skoruppa08}. The left hand side then collapses to a single term, $N:=\dim S_{3/2}(\Theta^{m+K})$, and we can write the dimension formula \eqref{eq:gannonRR} in a simpler form,
\begin{eqnarray}
    N = \frac{9d}{8}-\frac{\alpha_1}{2}-\frac{\beta_1+2\beta_2}{3}-\sum_{j=1}^d \Delta_j
    \label{eq:rrN}
\end{eqnarray}
Therefore, to compute $N$ for a given $m+K$, or equivalently a triple $(p_1,p_2,p_3)$, one needs only to construct the $S$ transformation matrix and count the eigenvalue multiplicities of $S$ and $\mu ST$. In Table \ref{tab:tab1}, we give the index $m$, representation dimension $d$, $N$, and the corresponding $c_{\rm eff}$ for a number of examples that were also studied in \cite{ACDGO2}, which match exactly.
\begin{table}[ht]
        \centering
            \begin{tabular}{|l|l|l|l|l|}
                \hline
                $(p_1,p_2,p_3)$ & $m$   & $d$ & $N$ & $c_{\rm eff}$ \\ \hline
                $(2,3,5)$       & $30$  & 2   & $1$ & $1/5$        \\ \hline
                $(2,3,7)$       & $42$  & 3   & $1$ & $1/7$        \\ \hline
                $(2,3,11)$      & $66$  & 5   & $2$ & $25/11$      \\ \hline
                $(2,5,7)$       & $70$  & 6   & $1$ & $3/35$       \\ \hline
                $(2,3,13)$      & $78$  & 6   & $1$ & $1/13$       \\ \hline
                $(2,3,17)$      & $102$ & 8   & $2$ & $25/17$      \\ \hline
                $(3,5,7)$       & $105$  & 12  & $2$ & $8/35$       \\ \hline
                $(2,5,11)$      & $110$ & 10  & $2$ & $27/55$      \\ \hline
                $(2,3,19)$      & $114$ & 9   & $2$ & $25/19$      \\ \hline
                $(2,7,13)$      & $182$ & 18  & $2$ & $27/91$      \\ \hline
            \end{tabular}
        \caption{Data for $c_{\rm eff}$ using \eqref{eq:rrN} is given for some Seifert manifolds $(p_1,p_2,p_3)$. These are the same examples given in Section 5.5 of \cite{ACDGO2}, up to a shift of $+1$ due to differing normalizations.}
        \label{tab:tab1}
\end{table}
With this confirmation that our prediction agrees with the data found via resurgent continuation, we make the following conjecture:
\begin{myconj}
    Let $\{\Delta_j\}$ be the ordered multiset of Chern-Simons values associated to the orientation reversed Brieskorn sphere $\overline{\Sigma(p_1,p_2,p_3)}$, and $N=\dim S_{3/2}(\Theta^{m+K})$. Then, $c_{\rm eff} = 24\Delta_N$.
\end{myconj}
Through some experimentation with this formula for $N$, we can study the behavior of the eigenvalue multiplicities $\alpha_i$ and $\beta_i$. An immediate observation that can be made is that the $\alpha_i$ and $\beta_i$ terms each scale linearly with $d$. In fact, it can be seen by experimenting over a large number of representations $\Theta^{m+K}$ that the formula can be greatly simplified.
\begin{eqnarray}
        \frac{\alpha_1}{2} = \frac{1}{2} \left\lfloor \frac{d}{2} \right\rfloor , \quad \frac{\beta_1+2\beta_2}{3} = \begin{cases} \frac{d}{3} \qquad {\rm if} \,\, 3  \big |  d \\
        \frac{d-1}{3} \,\,\,\,\,\, {\rm if} \,\, 3\nmid d
        \end{cases}
\end{eqnarray}
where $a|b$ means $a$ divides $b$, and $a \nmid b$ means $a$ does not divide $b$. This simplification removes most of the complexity when computing $N$, allowing us to probe the behavior of $c_{\rm eff} = 24\Delta_N$ as representations get extremely large. In Figure \ref{fig:allP}, we give a scatter plot of $c_{\rm eff}$ as a function of $m=p_1p_2p_3$, where we restrict to prime $p_i$.
This plot is quite striking, as it suggests that there is some universal behavior that forces the effective central charge to converge towards $c_{\rm eff}=1$ as the representations get large. Given the relationship between large index Brieskorn spheres and small surgeries on knot complements \cite{GJ23,GM19,ceffmod25,CDGG23}, this suggests that this feature may have some interesting interpretation in quantum topology.

\begin{figure}
    \centering
    \includegraphics[width=1\linewidth]{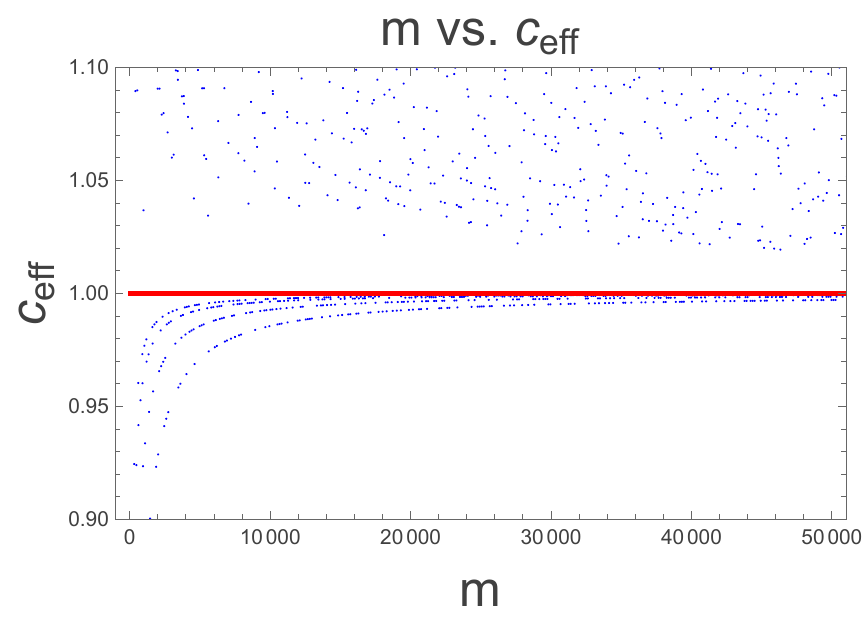}
    \caption{Values of $c_{\rm eff}$ in the class $\Sigma(2,3,p_3)$ as a function of $m=6p_3$ for prime $p_3$ (blue dots). A red line at $c_{\rm eff}=1$ is also plotted.}
    \label{fig:23p}
\end{figure}

Particularly interesting number theoretic features appear when we restrict the Seifert labels. One such class of manifolds are the Brieskorn spheres $\Sigma(2,3,p_3)$ for $p_3$ prime. In Figure \ref{fig:23p}, we plot $c_{\rm eff}$ for this class.
It can be immediately seen from the plot that there are ``bands" of points that asymptote to $c_{\rm eff} =1$. By analyzing the data $(p_3,c_{\rm eff})$, it becomes clear that $c_{\rm eff}$ behaves differently depending on the residue class $n$ of $p_3 \mod 24$ with $\gcd(n,24)=1$. For all $p_3 \equiv n_- \pmod{24}$, where $n_- \in \{1,5,7,13\}$, we find $c_{\rm eff}<1$, while for $p_3 \equiv n_+ \pmod{24}$ with $n_+ \in \{11,17,19,23\}$ we find $c_{\rm eff}>1$. Furthermore, for $c_{\rm eff}<1$ there are four bands, each of which correspond to a single $n_-$. Remarkably, when $p_3 \equiv n_- \pmod{24}$ we find an exact formula,
\begin{eqnarray}
    \hspace{-.4cm}c_{\rm eff}(2,3,p_3) \!=\! 1\!-\!\frac{\gcd (p_3-1,24)}{p_{3}}, \quad \!\! p_3 \equiv n_- \!\!\!\!\! \mod 24
\end{eqnarray}
This formula matches the predicted values for $c_{\rm eff}=24\Delta_N$ using \eqref{eq:rrN} for all triples $(2,3,p_3)$ with $m < 10^6$. While the examples with $p_3 \equiv n_+$ do not fit into as simple a formula as the $n_-$ class, there appears to be a more complicated underlying number theoretic structure that determines $c_{\rm eff}$.

We conclude with a comment on the case of Seifert manifolds with four exceptional fibers $\Sigma(p_1,p_2,p_3,p_4)$. The associated $\widehat{Z}$-invariants are combinations of weight $1/2$ and weight $3/2$ false theta functions \cite{Chung18}, and recently the weight $3/2$ contribution to the invariants of the orientation reversed manifold were shown to be related to the special mock Jacobi forms $\mathcal{Q}_m$ appearing in $\mathcal{N}=4$ quarter-BPS dyon counting \cite{DMZ12,AD26}. It has previously been shown that the mock modular forms $X^{m+K}$ appearing in the theta decomposition of these forms can always be chosen to be optimal when $m=p_1p_2p_3p_4$ \cite{DMZ12}, giving us another check for the formula \eqref{eq:rrN}. In the next section, which shows the individual terms in \eqref{eq:rrN} for some examples and how they sum to integers, we include the simplest four-fiber examples of $\Sigma(2,3,5,7)$ and $\Sigma(2,3,5,11)$.

In this final section we give explicit computations for $N$ for a few example Seifert manifolds with three and four exceptional fibers. The integrality of the right hand side of \eqref{eq:rrN} is established in \cite{Gannon13}, but it is instructive to see how the individual terms fit together. The main computational ingredient is the $\mathcal S$-transformation matrix, which is denoted by $S^{m+K}$, or just $S$ if the context is clear. We will present a general construction for $S^{m+K}$ for the case of $m=p_1p_2p_3$ and $K=\{p_1p_2,p_1p_3,p_2p_3\}$, i.e. the Seifert manifolds $\Sigma(p_1,p_2,p_3)$, and for each example will list the eigenvalue multiplicities, and show that the computation gives an integer. The construction for representations $\Theta^{m+K}$ for four fibers is conceptually similar, and is described in more detail in \cite{Hikami2}.

The matrices $S^{m+K}$ can be expressed as sums over the usual $S$ matrices for Weil representations of index $m$, which consist of terms of the form
\begin{eqnarray}
    S_{ab} = \sqrt{\frac{2}{m}} \sin\left(\frac{\pi a b}{m}\right)
\end{eqnarray}
The final expression for $S^{m+K}$ will be sums over these entries with particular arithmetic conditions on $a,b$. Define the matrix $\Omega_m(n)$ for $n$ that divide $m$ and with entries
\begin{eqnarray}
    \Omega_m(n)_{r,r'} = \delta_{r,-r' \!\!\!\!\pmod{2n}} \delta_{r,r' \!\!\!\! \pmod{2m/n}}
\end{eqnarray}
The Chern-Simons values can be determined by taking all integers $r_j$ from $1$ to $m=p_1p_2p_3$ such that the $p_i$ do not divide $r_j$ for $i=1,2,3$, and then selecting the minimal representatives $r_j$ of the class $r_j^2/4m \pmod{1}$:
\begin{eqnarray}
    \Delta_j = \frac{r_j^2}{4m} \pmod{1}
\end{eqnarray}
Next, let the operator $\Omega_{m,n}(r)$ be the unique value $r' \pmod{2m}$ such that $\Omega_m(n)_{r,r'}=1$. Then we form the orbits of the representatives $r_j$,
\begin{eqnarray}
    \mathcal{O}_j = \{r_j,\Omega_{m,p_1p_2}(r_j),\Omega_{m,p_1p_3}(r_j),\Omega_{m,p_2p_3}(r_j)\}
\end{eqnarray}
As a final ingredient, define
\begin{eqnarray}
    \chi_i(a) = \begin{cases}
        +1 , \quad a\in \mathcal{O}_i\\
        -1, \quad -a \in \mathcal{O}_i \\
        0, {\rm \,\,\,\,\,\,\,\,\,\, otherwise}
    \end{cases}, \quad a\in \mathbb{Z}/2m\mathbb{Z}
\end{eqnarray}
Finally we define the $S$ matrix for the $\Theta^{m+K}$ representations, 
\begin{eqnarray}
    S^{m+K}_{ij} &=& \sqrt{\frac{1}{8m}} \sum_{a \in \mathcal{O}_i}\sum_{b \in \mathcal{O}_{j}} \chi_i(a)\chi_j(b) \sin\left(\frac{ a  b\pi}{m}\right)
\end{eqnarray}
The $T$ transformation is implemented by the matrix 
\begin{eqnarray}
    T={\rm diag}\left(e^{2\pi i \Delta_1},e^{2\pi i \Delta_2},\dots,e^{2\pi i \Delta_d}\right)
\end{eqnarray}
We only list the explicit $S$ and $T$ matrices for the representation associated to the Poincar\'e homology sphere $\Sigma(2,3,5)$, which has $d=2$,
\begin{eqnarray}
    S=\sqrt{\frac{8}{30}}\left(
\begin{array}{cc}
 \sin \left(\frac{\pi }{30}\right)+\sin \left(\frac{19 \pi }{30}\right) & \sin \left(\frac{7 \pi }{30}\right)+\sin \left(\frac{17\pi }{30}\right) \\
 \sin \left(\frac{7 \pi }{30}\right)+\sin \left(\frac{17\pi }{30}\right) & -\sin \left(\frac{\pi }{30}\right)-\sin \left(\frac{19 \pi }{30}\right) \\
\end{array}
\right) \nonumber
\end{eqnarray}
and
\begin{eqnarray}
    T = {\rm diag}\left(e^{2\pi i \frac{1}{120}},e^{2\pi i \frac{49}{120}}\right)
\end{eqnarray}
Recall that $\alpha_n$ are the multiplicities of the eigenvalues $(-1)^n$ in $S$, while $\beta_n$ are the multiplicities of the eigenvalues $\zeta_3^n = e^{2\pi in/3}$ in the matrix $\mu ST$, where $\mu$ is a constant chosen such that $(\mu ST)^3=I$. We list the multiplicities as a vector $\vec \alpha=(\alpha_0,\alpha_1)$, and $\vec \beta=(\beta_0,\beta_1,\beta_2)$ below, along with the sum over $\Delta_j$
\begin{eqnarray}
    \vec \alpha = (1,1), \quad \vec \beta=(1,1,0), \quad \sum_{j=1}^2 \Delta_j = \frac{5}{12}
\end{eqnarray}
which gives
\begin{eqnarray}
    N = 2*\frac{9}{8} -\frac{1}{2} - \frac{1}{3} - \frac{5}{12} = 1
\end{eqnarray}
which gives the effective central charge $c_{\rm eff}=1/5$. The next simplest example is $(2,3,7)$, where we have $d=3$ and
\begin{eqnarray}
    \vec \alpha = (2,1), \quad \vec \beta=(1,1,1), \quad \sum_{j=1}^3 \Delta_j = \frac{7}{8}
\end{eqnarray}
giving
\begin{eqnarray}
    N=3*\frac{9}{8} - \frac{1}{2}-\frac{3}{3} - \frac{7}{8}=1
\end{eqnarray}
and $c_{\rm eff}=1/7$. For $(2,3,11)$, we have $d=5$ and
\begin{eqnarray}
    \vec \alpha = (3,2), \quad \vec \beta=(2,2,1), \quad \sum_{j=1}^5 \Delta_j = \frac{31}{24}
\end{eqnarray}
giving
\begin{eqnarray}
    N=5*\frac{9}{8}-\frac{2}{2}-\frac{4}{3}-\frac{31}{24}=2
\end{eqnarray}
This is the first example encountered that is non-optimal, and has $c_{\rm eff}=25/11$. For the last three-fiber example, $(2,3,13)$ with $d=6$, we return to optimality,
\begin{eqnarray}
    \vec \alpha = (3,3), \quad \vec \beta=(2,2,2), \quad \sum_{j=1}^6 \Delta_j = \frac{9}{4}
\end{eqnarray}
which gives
\begin{eqnarray}
    N=6*\frac{9}{8}-\frac{3}{2}-\frac{6}{3}-\frac{9}{4}=1
\end{eqnarray}
and $c_{\rm eff}=1/13$.

We finish with the four-fiber Seifert manifolds with labels $(2,3,5,7)$ and $(2,3,5,11)$. The dimension of the representations $\Theta^{m+K}$ is given by 
\begin{eqnarray}
    d=\frac{1}{8}\prod_{i=1}^4(p_i-1).
\end{eqnarray}
We use $k=1/2$ in \eqref{eq:gannonRR}, since the Bruinier-Funke pairing is now between the polar part of a weight $2-k=3/2$ mock modular form and a weight $k=1/2$ cusp form. However, we do not know of a similar vanishing theorem for the second term on the left hand side of \eqref{eq:gannonRR} as in the $k=3/2$ case, so we define $N:=\dim S_{1/2}(\Theta^{m+K})-\dim S_{3/2}(\overline{\Theta^{m+K}})$. Regardless, we find that the right hand side computes the expected value of $N=1$, coinciding with optimality as expected from \cite{DMZ12}. The rest of the computations are the same. We give the eigenvalue multiplicities $\alpha_n$, $\beta_n$, and the sum of the Chern-Simons values, starting with $(2,3,5,7)$ with $d=6$:
\begin{eqnarray}
    \vec \alpha = (3,3), \quad \vec \beta = (2,2,2), \quad \sum_{j=1}^6\Delta_j = \frac{7}{4}
\end{eqnarray}
which gives,
\begin{eqnarray}
    N=6*\frac{25}{24}-\frac{3}{2}-\frac{6}{3}-\frac{7}{4} = 1
\end{eqnarray}
and is consistent with the expected optimality of $(2,3,5,7)$. For $(2,3,5,11)$, which has $d=10$, we find
\begin{eqnarray}
    \vec \alpha = (5,5), \quad \vec \beta = (4,3,3), \quad \sum_{j=1}^{10}\Delta_j = \frac{47}{12}
\end{eqnarray}
which gives,
\begin{eqnarray}
    N=10*\frac{25}{24}-\frac{5}{2}-\frac{9}{3}-\frac{47}{12} = 1
\end{eqnarray}
and is optimal as expected. We have computed $N$ for Seifert labels $(p_1,p_2,p_3,p_4)$ such that $m=p_1p_2p_3p_4<10^5$, all of which give $N=1$.

We conclude this letter with some remarks on future directions. Seifert homology spheres are among the simplest three manifolds, and it would be interesting to see how this method extends to non-homology spheres, or even non-Seifert manifolds. There exists a form of modularity for the $\widehat{Z}$ invariants associated to these manifolds, and there may be a similar way of computing $c_{\rm eff}$. It would also be interesting to see if the genus zero property for the representations admitting optimal $q$-series can be directly related to the condition of $N=1$, which signals optimality in our approach. Lastly, the physical interpretation of $c_{\rm eff} \to 1$ for large Seifert labels $(p_1,p_2,p_3)$ may have interesting implications for the $T[M_3]$ theories, as well as for the behavior of $\widehat{Z}$ invariants for small surgeries on knot complements.

\noindent{\bf Acknowledgement:} This material is based upon work supported by the U.S. Department of Energy, Office of Science, Office of High Energy Physics under Award Number DE-SC0010339. The author thanks G. Dunne, O. Costin, and A. Saraeb for helpful discussions of the content in this paper. The author also extends a special thanks to O. {\"O}ner for early discussions that inspired this work.

\end{document}